\documentclass[journal]{IEEEtran}
\usepackage[colorlinks,urlcolor=blue,linkcolor=blue,citecolor=blue]{hyperref}
\usepackage{amsmath}
\usepackage{multirow}
\usepackage{wrapfig}
\usepackage{pifont}
\usepackage{graphicx}
\usepackage{colortbl}
\usepackage[table]{xcolor}
\usepackage{amssymb}
\usepackage{booktabs}
\usepackage{balance}
\begin{document}

\title{REVISION: A Roadmap on Adaptive Video Streaming Optimization}

\author{Farzad Tashtarian, Christian Timmerer\\
Christian Doppler Laboratory ATHENA, Alpen-Adria Universität Klagenfurt, Klagenfurt, Austria
}

\newcommand{\etal}{\textit{et al. }}
\newcommand{\ie}{\emph{i.e.}, }
\newcommand{\revi}{REVISION}
\newcommand{\eg}{\emph{e.g.}, }
\newcommand{\etc}{\emph{etc.\xspace}}
\maketitle

\begin{abstract}
Due to the soaring popularity of video applications and the consequent rise in video traffic on the Internet, technologies like HTTP Adaptive Streaming (HAS) are crucial for delivering high Quality of Experience (QoE) to consumers. HAS technology enables video players on consumer devices to enhance viewer engagement by dynamically adapting video content quality based on network conditions. This is especially relevant for consumer electronics as it ensures an optimized viewing experience across a variety of devices, from smartphones to smart TVs.
This paper introduces REVISION, an efficient roadmap designed to enhance adaptive video streaming—a core feature of modern consumer electronics. The REVISION optimization triangle highlights three essential aspects for improving streaming: Objective, Input Space, and Action Domain. Additionally, REVISION proposes a novel layer-based architecture tailored to refine video streaming systems, comprising Application, Control and Management, and Resource layers. Each layer is designed to optimize different components of the streaming process, which is directly linked to the performance and efficiency of consumer devices.
By adopting the principles of the REVISION, manufacturers and developers can significantly improve the streaming capabilities of consumer electronics, thereby enriching the consumer's multimedia experience and accommodating the increasing demand for high-quality, real-time video content. This approach addresses the complexities of today's diverse video streaming ecosystem and paves the way for future advancements in consumer technology.\end{abstract}

\IEEEpeerreviewmaketitle
\section{Introduction}
\begin{figure}[t]
  \begin{center}
    \includegraphics[width=0.5\textwidth]{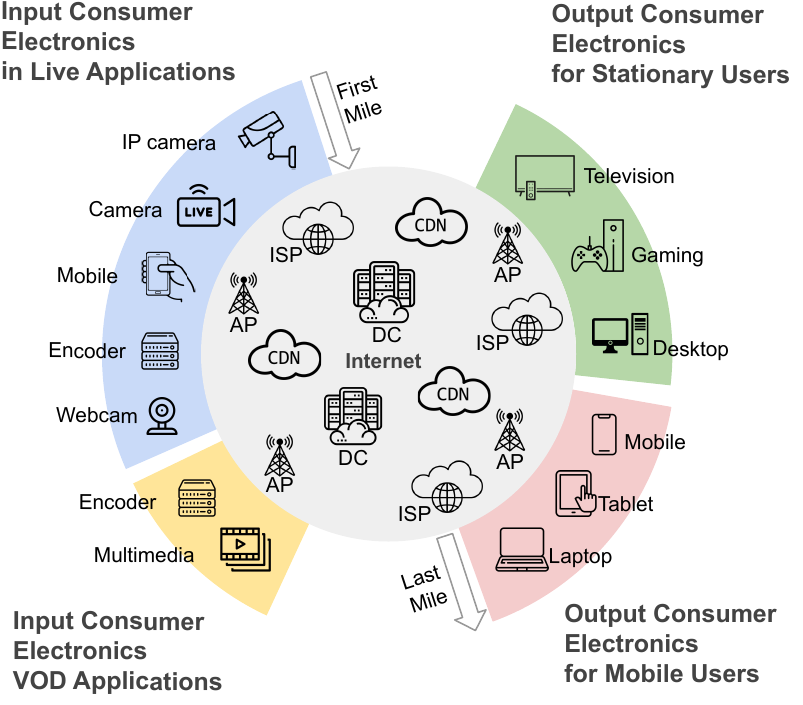}
  \end{center}
  \caption{Video streaming applications.}
  \label{fig:app}
\end{figure}

have seen remarkable growth, significantly influenced by a 24\% increase in Internet video traffic during 2022~\footnote{Sandvine Inc. "The Global Internet Phenomena Report", \url{https://www.sandvine.com/phenomena}}. This surge highlights the escalating demand for technologies such as HTTP Adaptive Streaming (HAS), which is now pivotal in managing video traffic and ensuring high Quality of Experience (QoE) for consumers of online video content. The evolution of video streaming technologies, including advanced video compression algorithms~\cite{nightingale2014video,hamidouche2022versatile}, has enabled seamless content delivery across diverse consumer electronics such as smartphones, smart TVs, laptops, and tablets. These devices are at the heart of how video streaming, including Video On Demand (VOD) and live streaming, transforms consumer interactions with digital content.
VOD services have revolutionized access to extensive content libraries, allowing consumers to customize their viewing experiences~\cite{khani2023recl}, while live streaming has enabled real-time engagement with events such as sports, interactive social media sessions, and online gaming~\cite{mattioli2020history}, thereby connecting global audiences in unprecedented ways~\cite{chen2023understanding}. The transition from less than 1\% to nearly 18\% of Internet traffic being attributed to live streaming from 2015 to 2022 underscores its significant impact on consumer technology~\cite{cisco2020cisco}.
Figure~\ref{fig:app} illustrates the initial stage (first mile) of the streaming process, showcasing input devices and sources critical for both VOD and live applications. These sources send content across the Internet through various channels such as Data Centers (DCs), Content Delivery Networks (CDNs), Internet Service Providers (ISPs), and Access Points (APs). In the final stage (last mile) of streaming, a diverse array of consumer devices, which vary widely in hardware capabilities, connect and stream content, reflecting global trends such as the preference for smartphones in markets like China for video consumption and social networking~\cite{noor2015connected}.
\begin{figure*}[t]
    \centering
    \includegraphics[width=1\linewidth]{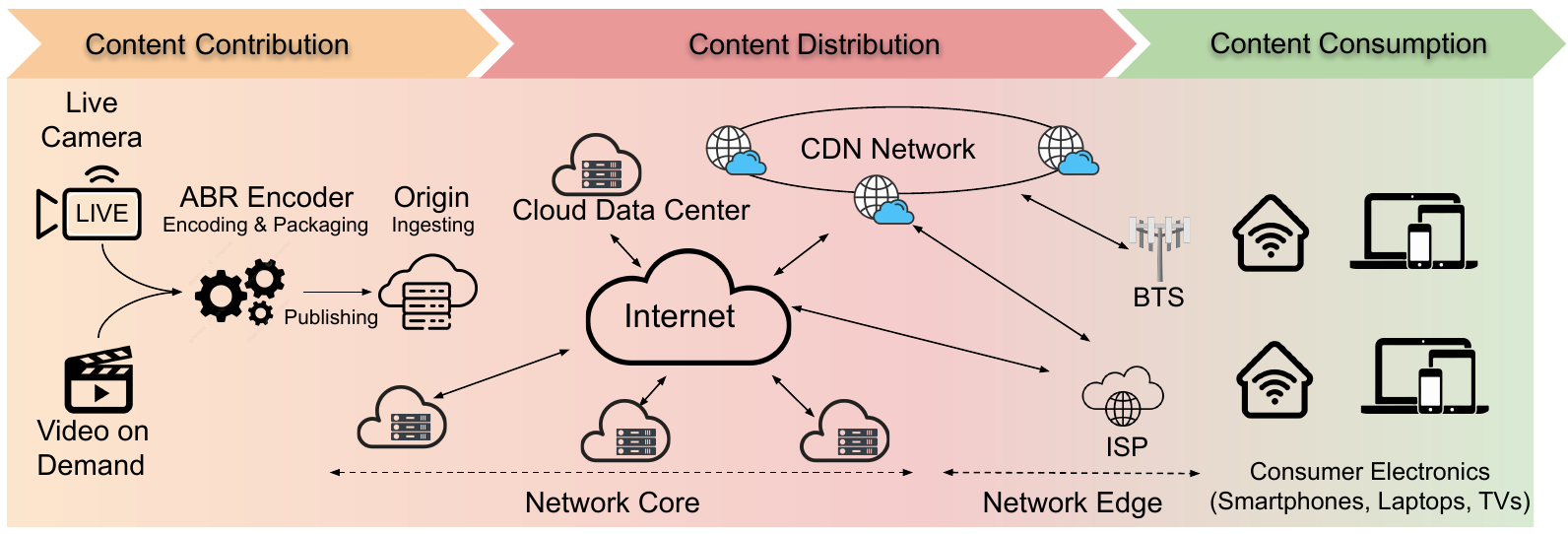}
    \caption{An adaptive video streaming pipeline.}
    \label{fig:end-to-end}
\end{figure*}
HTTP Adaptive Streaming (HAS) enhances this consumer experience by segmenting videos into multiple quality representations, allowing devices to select the optimal video quality dynamically based on network conditions~\cite{bentaleb2017sdnhas}. This adaptive capability, facilitated by bitrate ladders and adaptive bitrate (ABR) algorithms described in clients' manifest files, is crucial for maintaining excellent QoE, which includes considerations of video quality, stall duration, and stability. The increasing consumer expectation for high-quality VOD and live streaming services makes the study of QoE a critical aspect of consumer electronics.
\subsection{Video Streaming Pipeline}
Before addressing the problem of optimizing video streaming, we first investigate the intricacies of the video streaming pipeline and describe the main challenges and operations within each stage. As shown in Figure~\ref{fig:end-to-end}, the streaming pipeline can be divided into three stages~\cite{tashtarian2022hxl3}: \textit{Content Contribution}, \textit{Content Distribution}, and \textit{Content Consumption}.
\begin{figure*}[t]
    \centering
    \includegraphics[width=1\linewidth]{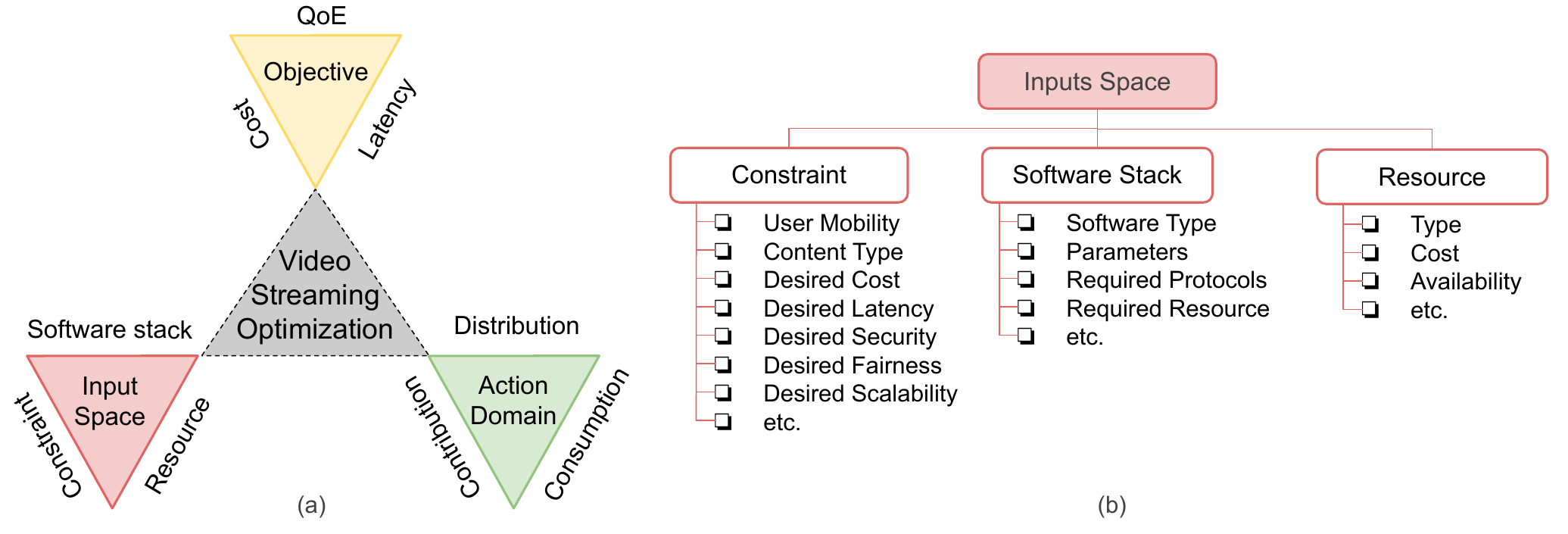}
    \caption{Optimization in video streaming: (a) REVISION optimization triangle and (b) Input Space.}
    \label{fig:input}
\end{figure*}
\begin{itemize}
    \item \textbf{Content Contribution}. This is the first mile of the video streaming pipeline and comprises three main entities: ($i$) multimedia source:  in general, based on the type of application, this can be a live camera source capturing a live stream or recorded video in the VOD application, ($ii$) adaptive bitrate (ABR) encoder: deployed either on-premise or in the cloud for media encoding, packaging, and encryption using digital rights management (DRM) services, crucial for protecting content accessed on consumer devices, and ($iii$) origin server: responsible for ingesting and distributing packaged media content into the network, ensuring content availability and security for consumer access.
    \item \textbf{Content Distribution}. The main responsibility of this stage is to disseminate the generated content from the origin server to the HAS players. This involves Cloud Data Centers and CDNs at the network core, extending to the edge network by leveraging base stations (BSs) and Internet service providers (ISPs). This stage is crucial for enhancing the performance of consumer devices as it ensures low latency and high-quality streaming through efficient network management and edge computing capabilities. Once the video content is transferred to the CDN by the origin server~\footnote{Note that a CDN may already contain the origin server.}, the data is stored for faster access to meet HAS players' requests, and computational resources at the network's edge are utilized for functions like transcoding, further reducing latency on consumer devices.
    \item \textbf{Content Consumption}. The last mile of the video streaming pipeline involves the content consumption stage comprising multiple heterogeneous consumer electronics like smartphones and smart TVs that support various HAS-based formats (\ie DASH, HLS, CMAF). Video streaming platforms often employ ABR techniques to dynamically adjust the quality of the video stream based on the viewer's network conditions. This stage directly impacts consumer technology by ensuring smooth playback on consumer devices even under varying bandwidth situations, thereby significantly improving the user experience across a wide range of consumer electronics.
\end{itemize}
\subsection{Primary Challenges in Video Streaming}
Enhancing the Quality of Experience (QoE) in video streaming, reducing latency, and managing costs are primary challenges in the rapidly evolving domain of digital media. As consumer expectations rise with technological advancements, streaming services strive to deliver high-quality video seamlessly across diverse network conditions and devices. \par \textbf{Quality of Experience}—Enhancing the QoE in video streaming is an ongoing challenge, as viewers' expectations continually rise aligned with technological advancements~\cite{peroni2023empowerment}. One of the primary challenges lies in optimizing video quality while balancing the constraints of varying network conditions and device capabilities, particularly in consumer electronics such as smartphones, tablets, and smart TVs. Since improving QoE involves addressing various factors, such as buffering interruptions, abrupt quality shifts, and stall duration during adaptive streaming, the challenge of consistently delivering high-quality streaming across a diverse range of consumer devices, resolutions, bitrates, and internet connections remains complex~\cite{peroni2023empowerment,huang2019comyco}.\par
\textbf{Latency}—
Latency is another critical challenge, particularly in live streaming scenarios, where real-time interactivity is preferable~\cite{lyko2024improving,zhang2021performance}. Latency in video streaming refers to the delay between the moment a video is captured or encoded and when it is displayed on the viewer's screen. Achieving low-latency streaming without compromising video quality and playback stability has been realized as a challenging issue and is particularly significant in consumer technology, where the immediacy of content delivery enhances user engagement and satisfaction.\par
\textbf{Cost}—Finally, as streaming services grow in popularity, scalability and efficient resource allocation become key challenges. These factors have a significant impact on the cost of running a streaming service, especially when accommodating millions of concurrent viewers accessing the same content. Effective cost management directly affects the affordability and accessibility of streaming services on consumer devices, making it a critical area of focus for sustaining the growth of consumer-oriented technologies.\par
\textit{Thus, improving QoE and latency in video streaming with reasonable cost can be defined as multifaceted challenges in both industry and academia. Many researchers have tackled the issue, endeavoring to enhance various aspects of QoE, latency, and cost.} These challenges are particularly significant in the consumer electronics sector, where enhancing video streaming quality directly translates to improved user satisfaction and broader adoption of streaming technologies across devices. However, the absence of a dedicated model for video streaming systems remains apparent, despite some proposed models making use of Software-Defined Networks (SDN)~\cite{tashtarian2018s2vc,bentaleb2017sdnhas,erfanian2020optimizing} or 
those that enable reuse of infrastructure to deliver scalable, virtualized, distributed, federated services~\cite{iraschko2019next}. In this paper, to address the issues and challenges in video streaming, we first present REVISION as a roadmap for optimizing adaptive video streaming. As the name implies, REVISION stands for a \textit{R}oadmap on adaptiv\textit{E} \textit{VI}deo \textit{S}treaming optimization\textit{ION}. We define the REVISION optimization triangle, which includes the following elements: \textit{Objective}, \textit{Input Space}, and \textit{Action Domain}, corresponding to the three pivotal questions on optimizing video streaming, respectively: (i) Which metrics should be optimized? (ii) Which parameters should be considered? (iii) Which elements and parameters should be tuned? This structured approach is vital for developing consumer devices that can adaptively manage streaming quality in real-time, crucial for enhancing the end-user's viewing experience on consumer electronics. We then analyze several cutting-edge approaches regarding the REVISION optimization triangle. We finally present our layer-based REVISION architecture consisting of three distinct layers: the \textit{Application} layer, the \textit{Control and Management} layer, and the \textit{Resource} layer. Within each layer, we define the key components and modules and their responsibilities. 
The proposed architecture is a comprehensive model for optimizing video streaming, offering a structured approach to improving the streaming experience and directly impacting the performance and efficiency of consumer electronics that utilize streaming services.\\
\section{Optimizing Video Streaming with REVISION}
Optimizing video streaming presents significant challenges when considering the various tasks and devices involved in the streaming pipeline. The solutions should support an end-to-end perspective, as local optimization in each pipeline stage may not guarantee improvement of the target objective in a real streaming scenario~\cite{rahman2023content,tashtarian_artemis_2023}.
Therefore, any optimization task in video streaming involves the selection of a suitable objective function and the subsequent design of an efficient approach. Figure~\ref{fig:input}a shows the REVISION video streaming optimization triangle that involves the determination of three main components: \textit{Objectives}, \textit{Input Space}, and \textit{Action Domain}. Each of these components comprises various parameters, which are crucial in enhancing the streaming experience on consumer technology platforms, directly impacting user satisfaction and device performance. In the following, we provide a detailed introduction to these components.\par
\begin{figure*}[t]
    \centering
    \includegraphics[width=1\linewidth]{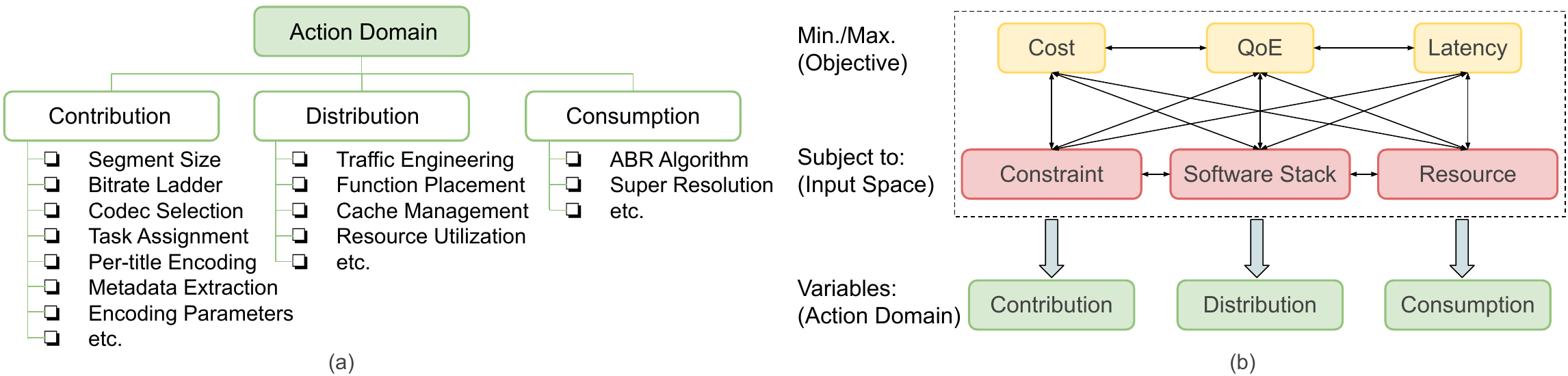}
    \caption{‌(a) Action Domain and (b) optimization form of REVISION.}
    \label{fig:action}
\end{figure*}
\subsection{Objective}
In video streaming, improving clients' perceived QoE is one of the primary objectives. Enhancing QoE not only attracts more customers and increases revenue for video streaming providers but also significantly improves the usability and functionality of consumer electronics that rely on streaming content. While many QoE models consider factors such as stalls, bitrates, bitrate switches (smoothness), and startup delays, they often overlook a critical element: latency~\cite{peroni2023empowerment}.
Latency, in the context of real-time video streaming, refers to the delay or lag between the moment a video is captured, transmitted, and eventually displayed on a viewer's screen~\cite{tashtarian2022hxl3}. Thus, latency is an essential concern, particularly for real-time video streaming applications, and it is pivotal for enhancing the interactive experience on consumer devices, making it worth considering as an objective metric.
Theoretically, achieving the best possible QoE and latency values would be obtained by allocating a maximum of resources. However, this approach is impractical in the real world. Hence, some studies focus on optimizing the overall associated resource costs, including computation, bandwidth, and storage~\cite{tashtarian_lalisa_2023,tashtarian_artemis_2023,erfanian2021lwte}. Their goal is to reduce these costs without significantly compromising QoE and/or latency, which are critical for maintaining high performance and low operational costs in consumer technology. Therefore, any combination of QoE, latency, and cost can be selected as the objective function in video streaming optimization research while adjusting the coefficient values as needed~\cite{tashtarian_artemis_2023,erfanian2021oscar}. \par

\subsection{Input Space}
The input space encompasses all parameters and variables that should be considered for optimizing the video streaming pipeline, regardless of the selected objective function. For ease of explanation, we categorize the input space into three main categories: \textit{Constraint}, \textit{Software Stack}, and \textit{Resource} as shown in Figure~\ref{fig:input}b.
The constraint category comprises parameters linked to the application type and the given constraints by the system administrator\footnote{A system administrator in a video streaming system is responsible for the overall management, configuration, and maintenance of the technical infrastructure that facilitates the delivery of video content to users.} and significantly influences the solution design, with direct implications for the performance and capabilities of consumer electronics. For example, the encoding parameters can be fine-tuned based on factors like~\textit{Content Type}, which may include categories such as animation, action, documentary, and sport. Additionally, essential considerations, \eg \textit{desired latency},~\textit{cost}, are realized as important constraints during the design process of approaches tailored to consumer technology needs. To launch a video streaming application, a series of protocols, software components, and functions should be employed at each stage of the streaming pipeline. Therefore, the second group of parameters within the input space, known as the~\textit{Software Stack}, encompasses various properties, including \textit{Type} that refers to the type of application that can be utilized in the pipeline such as encoders, transcoders, web servers, and media players, along with their associated parameters. It is also important to consider the required protocols and resources for utilizing each component within this category. 
The last group in the input space is \textit{Resource} which consists of critical parameters related to various resource types, \ie computation, storage, and bandwidth. For instance, \textit{Cost} and \textit{Availability} refer to the financial implications and amount of available resources that must be taken into account when running different video streaming applications.\par

\subsection{Action Domain}
The Action Domain is defined as the functions or variables in the video streaming pipeline that we intend to manipulate to optimize the objective function. As shown in Figure~\ref{fig:action}a, the Action Domain can be divided into three parts according to the video streaming pipeline stages: contribution, distribution, and consumption stages, where each of them encompasses different items.
For instance, in the contribution stage, we have a set of functions and variables related to the following processes: gathering content from input sources, encoding, packaging, and eventually sending it into the distribution network. In the distribution and consumption stages, there are some techniques and methods that can be manipulated and tuned to optimize the video streaming performance. For example, in some studies~\cite{erfanian2021oscar,erfanian2021lwte-live}, researchers attempt to optimize the objectives by applying novel video traffic engineering, placing lightweight functions at the network edge (\eg transcoding), tuning cache management policies, etc. Furthermore, some studies focus on the ABR algorithm at the client side to optimize various objective functions~\cite{nguyen2023performance,tashtarian_lalisa_2023}. This optimization is particularly vital for consumer devices, where efficient use of bandwidth and enhanced video quality are critical for device performance and user satisfaction. Moreover, implementing super-resolution at the client side enables consumer devices to enhance the visual quality of low-resolution content directly on their devices, thus reducing the burden on network bandwidth and server resources and improving the overall streaming experience for the consumer. \par
\begin{table*}[t]
\definecolor{c1}{HTML}{fff2ccff}
\definecolor{c2}{HTML}{f4ccccff}
\definecolor{c3}{HTML}{d9ead3ff}
\centering
\caption{Objectives, Input Space, and Action Domain of recent studies.}
\label{tbl1}
\resizebox{\textwidth}{!}{%
\begin{tabular}{@{}|c||ccc||ccc||ccc|@{}}
\hline
\multirow{2}{*}{\textbf{Study}}&
  \multicolumn{3}{c||}{\cellcolor{c1}\textbf{Objective}} &
  \multicolumn{3}{c||}{\cellcolor{c2}\textbf{Input Space}} &
  \multicolumn{3}{c|}{\cellcolor{c3}\textbf{Action Domain}} \\ 
 &
  \multicolumn{1}{c|}{\cellcolor{c1}\textit{Cost}} &
  \multicolumn{1}{c|}{\cellcolor{c1}\textit{QoE}} &
  \cellcolor{c1}\textit{Latency} &
  \multicolumn{1}{c|}{\cellcolor{c2}\textit{Constraint}} &
  \multicolumn{1}{c|}{\cellcolor{c2}\textit{Software Stack}} &
  \cellcolor{c2}\textit{Resource} &
  \multicolumn{1}{c|}{\cellcolor{c3}\textit{Contribution}} &
  \multicolumn{1}{c|}{\cellcolor{c3}\textit{Distribution}} &
  \cellcolor{c3}\textit{Consumption} \\ \hline\hline
LwTE~\cite{erfanian2021lwte} &
  \multicolumn{1}{c|}{ \checkmark} &
  \multicolumn{1}{c|}{ } &
    &
  \multicolumn{1}{c|}{} &
  \multicolumn{1}{c|}{Transcoder} &
  \begin{tabular}[c]{@{}c@{}}Storage \\ Bandwidth\end{tabular} &
  \multicolumn{1}{c|}{Metadata Ext$.^*$} &
  \multicolumn{1}{c|}{} &
   \\ \hline
HxL3~\cite{tashtarian2022hxl3} &
  \multicolumn{1}{c|}{ \checkmark} &
  \multicolumn{1}{c|}{ } &
    &
  \multicolumn{1}{c|}{Desi. Latency} &
  \multicolumn{1}{c|}{\begin{tabular}[c]{@{}c@{}}Transcoder\\ VRP\end{tabular}} &
  \begin{tabular}[c]{@{}c@{}}Computation\\ Bandwidth\end{tabular} &
  \multicolumn{1}{c|}{} &
  \multicolumn{1}{c|}{\begin{tabular}[c]{@{}c@{}}Traffic Eng.\\ Resource Uti.\end{tabular}} &
   \\ \hline
Pensieve~\cite{mao2017neural} &
\multicolumn{1}{c|}{ } &
\multicolumn{1}{c|}{ \checkmark} &
 &
\multicolumn{1}{c|}{Desi. Scalability} &
\multicolumn{1}{c|}{\begin{tabular}[c]{@{}c@{}}Player\end{tabular}} &
\begin{tabular}[c]{@{}c@{}}Bandwidth\\Storage\end{tabular} &
\multicolumn{1}{c|}{} &
\multicolumn{1}{c|}{} &
ABR Alg. \\ \hline
SDNDASH~\cite{bentaleb2017sdnhas} &
\multicolumn{1}{c|}{ } &
\multicolumn{1}{c|}{ \checkmark} &
  &
\multicolumn{1}{c|}{\begin{tabular}[c]{@{}c@{}}Content Type\\ Desi. Latency\\Desi. Fairness\end{tabular}} &
\multicolumn{1}{c|}{\begin{tabular}[c]{@{}c@{}}Encoder\\ Player\\SND Controller\end{tabular}} &
\begin{tabular}[c]{@{}c@{}}Computation\\ Bandwidth\\Storage\end{tabular} &
\multicolumn{1}{c|}{} &
  \multicolumn{1}{c|}{\begin{tabular}[c]{@{}c@{}}Traffic Eng.\\ Resource Uti.\end{tabular}} &
ABR Alg. \\ \hline
H2BR~\cite{nguyen2023performance} &
  \multicolumn{1}{c|}{ } &
  \multicolumn{1}{c|}{ \checkmark} &
    &
  \multicolumn{1}{c|}{} &
  \multicolumn{1}{c|}{\begin{tabular}[c]{@{}c@{}}Player\\ Protocol\end{tabular}} &
  Bandwidth &
  \multicolumn{1}{c|}{} &
  \multicolumn{1}{c|}{} &
  ABR Alg. \\ \hline
LALISA~\cite{tashtarian_lalisa_2023} &
  \multicolumn{1}{c|}{ \checkmark} &
  \multicolumn{1}{c|}{ \checkmark} &
    &
  \multicolumn{1}{c|}{Content Type} &
  \multicolumn{1}{c|}{\begin{tabular}[c]{@{}c@{}}Encoder\\ Player\end{tabular}} &
  \begin{tabular}[c]{@{}c@{}}Computation\\ Bandwidth\end{tabular} &
  \multicolumn{1}{c|}{Bitrate Ladder} &
  \multicolumn{1}{c|}{} &
  ABR Alg. \\ \hline
ARTEMIS~\cite{tashtarian_artemis_2023} &
  \multicolumn{1}{c|}{ \checkmark} &
  \multicolumn{1}{c|}{ \checkmark} &
    &
  \multicolumn{1}{c|}{\begin{tabular}[c]{@{}c@{}}Content Type\\ Desi. Scalibility\end{tabular}} &
  \multicolumn{1}{c|}{\begin{tabular}[c]{@{}c@{}}Encoder\\ Player\end{tabular}} &
  \begin{tabular}[c]{@{}c@{}}Computation\\ Bandwidth\end{tabular} &
  \multicolumn{1}{c|}{Bitrate Ladder} &
  \multicolumn{1}{c|}{} &
   \\ \hline
OTEC~\cite{afzal2022otec} &
  \multicolumn{1}{c|}{ \checkmark} &
  \multicolumn{1}{c|}{ } &
   \checkmark &
  \multicolumn{1}{c|}{} &
  \multicolumn{1}{c|}{Transcoder} &
  \begin{tabular}[c]{@{}c@{}}Computation\\ Bandwidth\end{tabular} &
  \multicolumn{1}{c|}{Task Assingment} &
  \multicolumn{1}{c|}{} &
   \\ \hline
Bhat\etal\cite{bhat2021combining}&
  \multicolumn{1}{c|}{ } &
  \multicolumn{1}{c|}{ \checkmark} &
   &
  \multicolumn{1}{c|}{} &
  \multicolumn{1}{c|}{\begin{tabular}[c]{@{}c@{}}Encoder\\ Player\end{tabular}} &
  \begin{tabular}[c]{@{}c@{}}Computation\end{tabular} &
  \multicolumn{1}{c|}{Bitrate Ladder} &
  \multicolumn{1}{c|}{} &
   \\ \hline
SEGUE\cite{licciardello2022prepare}&
  \multicolumn{1}{c|}{ } &
  \multicolumn{1}{c|}{ \checkmark} &
   &
  \multicolumn{1}{c|}{Content Type} &
  \multicolumn{1}{c|}{Encoder} &
  \begin{tabular}[c]{@{}c@{}}Computation\\Bandwidth\end{tabular} &
  \multicolumn{1}{c|}{Bitrate Ladder} &
  \multicolumn{1}{c|}{Bandwidth Uti.} &ABR Alg.
   \\ \hline
GreenABR~\cite{turkkan2022greenabr}&
  \multicolumn{1}{c|}{\checkmark } &
  \multicolumn{1}{c|}{ \checkmark} &
   &
  \multicolumn{1}{c|}{} &
  \multicolumn{1}{c|}{Player} &
  \begin{tabular}[c]{@{}c@{}}Computation\\Bandwidth\\Energy\end{tabular} &
  \multicolumn{1}{c|}{} &
  \multicolumn{1}{c|}{} &ABR Alg.
   \\ \hline
MEDUSA~\cite{lorenzi2024medusa}&
  \multicolumn{1}{c|}{ } &
  \multicolumn{1}{c|}{ \checkmark} &
   &
  \multicolumn{1}{c|}{Desi. Scalibility} &
  \multicolumn{1}{c|}{\begin{tabular}[c]{@{}c@{}}Player\\ Encoder\end{tabular}} &  
  \begin{tabular}[c]{@{}c@{}}Bandwidth\end{tabular} &
  \multicolumn{1}{c|}{\begin{tabular}[c]{@{}c@{}}Codec Selection\\ Metadata Extraction\end{tabular}} &
  \multicolumn{1}{c|}{Bandwidth Uti.} &ABR Alg.
   \\ \hline
\end{tabular}}
\begin{flushleft}
*Abbreviations are "Alg.": Algorithm, "Desi.": Desired, "Ext.": Extraction, "Par.": Parameters., "Eng.": Engineering, "Uti.": Utilization, "Pla.": Placement, and "VRP": Virtual Reverse Proxy.
\end{flushleft}
\end{table*}

Table~\ref{tbl1} shows several recent studies' objectives, input space, and action domain. In the following, we investigate some of them. LwTE~\cite{erfanian2021lwte} and HxL3~\cite{tashtarian2022hxl3} focus on minimizing streaming costs. They achieve this by considering certain streaming functions as the input space and influencing the streaming pipeline through metadata extraction at the contribution stage or orchestrating streaming traffic at the distribution stage.
Other works, such as Pensieve~\cite{mao2017neural}, SDNDASH~\cite{bentaleb2017sdnhas}, and H2BR~\cite{nguyen2023performance}, aim to improve QoE by modifying/developing new ABR algorithms or optimizing network traffic using various entities from the input space.
In our recent research, LALISA~\cite{tashtarian_lalisa_2023} and ARTEMIS~\cite{tashtarian_artemis_2023}, we consider both cost and QoE as the multi-objective functions. We strive to optimize these functions by selecting an appropriate bitrate ladder. However, the distinction between LALISA and ARTEMIS, as shown in Table~\ref{tbl1}, lies in ARTEMIS's consideration of scalability as a constraint in the input space. This feature makes ARTEMIS agnostic to the ABR algorithm.

\subsection{Optimization Form of REVISION}
How does REVISION assist us in modeling the problem of optimizing video streaming? Figure~\ref{fig:action}b depicts an optimization model in the context of REVISION with the objective of optimizing (\ie maximizing or minimizing) a combination of QoE, cost, and latency. The input space can include all parameters and constraints provided by the system administrator (\eg desired latency from the \textit{Constraint} group) or the environment (\eg bandwidth fluctuations from the \textit{Resource} group). The action domain refers to the components that should be tuned by optimizing associated variables in the model.

The dependency graph among the objective and input space components reveals a strict influence among them. This dependency exists not only between the objective and input space but also among their components. This means that QoE, for example, impacts both cost and latency while being related to all components of the input space that are integral to enhancing the streaming experience on consumer devices. Therefore, it is essential to design an objective function with appropriate coefficient values for each item within the objective function to optimize performance and efficiency in consumer technology.
In some studies, QoE is selected as the sole objective~\cite{nguyen2023performance}; however, there are some constraints on resources that limit the total cost. In other words, the aforementioned objective values can be utilized in the model either directly in the objective function or implicitly by defining a couple of constraints~\cite{erfanian2021oscar}.

Another vital consideration is which parameters from the input space we should select when targeting different objective functions and specific action domains. Considering a high number of input space parameters may be particularly challenging for consumer electronics, where device constraints necessitate optimized performance with minimal overhead. 
It is worth mentioning that there is a tradeoff between the breadth of the input space and the quality of the solution. This means that as we increase the number of parameters in the optimization process, the results become more accurate and realistic; however, adding more parameters increases the complexity of the problem.
Therefore, we should find an efficient way to squeeze the input space while reaching realistic solutions that align with the capabilities and limitations of contemporary consumer electronics.  
Although the variables associated with the chosen action domain should be reflected in the input space, distinguishing input space parameters concerning the objective function becomes challenging due to strong correlations among various objective metrics and the elements of the input space. However, to narrow down the input space, we propose some alternative options as follows. 

The first approach is taking some realistic assumptions into account regarding the type of video streaming application. For example, if we aim to manipulate some procedures in the content contribution stage, \eg optimizing the bitrate ladder, making some assumptions about the content distribution and consumption can help us to reduce the number of input space parameters. For instance, in~\cite{tashtarian_lalisa_2023}, we tackle the problem of optimizing the bitrate ladder in live streaming by proposing realistic assumptions for both the distribution and consumption stages as follows: ($i$) fixed segment duration and ($ii$) negligible required storage at CDN for live content. These assumptions are particularly relevant for consumer electronics where efficiency and resource management are critical.
The second solution is to split the main problem into sub-problems and then sequentially solve them. This approach reduces the quality of results but is very practical~\cite{erfanian2021oscar}. Moreover, utilizing machine learning techniques to find the correlation between the input space and the variables in the action domain is an efficient way to reduce the complexity of the problem by selecting those parameters of the input space with higher correlations~\cite{tashtarian_artemis_2023}.\\

\section{REVISION Architecture}\par
\begin{figure*}[t]
    \centering
    \includegraphics[width=1\linewidth]{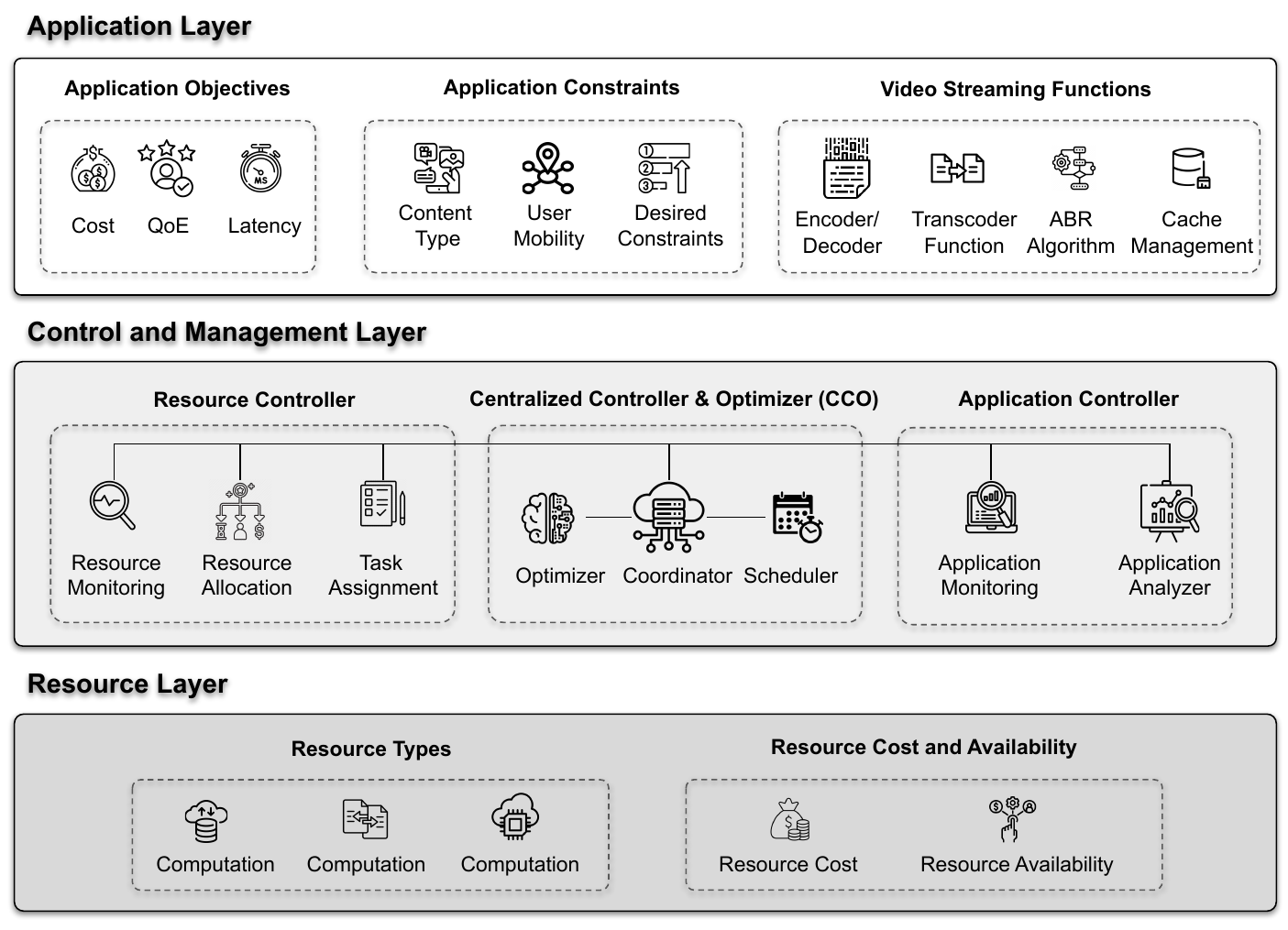}
    \caption{REVISION architecture for adaptive video streaming applications.}
    \label{fig:arch}
\end{figure*}

Regarding the components proposed for optimizing video streaming in Figure~\ref{fig:input}a,  a model to guide us in developing solutions is needed. The model should simplify the optimization solution's design process by highlighting the key modules and their correlations. Therefore, in this section, we propose the REVISION layer-based architecture for video streaming illustrated in Figure~\ref{fig:arch}. The model consists of three main layers: ($i$) \textit{Application}, ($ii$) \textit{Control and Management}, and ($iii$) \textit{Resource} layer. In the following, each layer is explained in detail, emphasizing their implementation in consumer electronics such as smart TVs, gaming consoles, and mobile devices to ensure high-quality streaming experiences.
\subsection{Application Layer} The application layer introduces three main blocks covering the objective items and some parts of the Input Space, denoted by \textit{Application Objectives}, \textit{Application Constraints}, and \textit{Video Streaming Functions} (see Application layer in Figure~\ref{fig:arch}). The Application Objective block encompasses all introduced objective items. Similarly, the Application Constraints block follows the items introduced in the Constraint component of the Input Space (see Figure~\ref{fig:input}), where the \textit{Desired Constraints} refers to a set of desired attributes given by the system administrator, \eg desired latency, cost, fairness, scalability, and security. This layer is particularly relevant to consumer electronics manufacturers, as it ensures devices can deliver tailored video streaming experiences that adapt to varying consumer demands and network environments. The last block, Video Streaming Functions, encompasses important video streaming functions and algorithms, \eg encoder/decoder~\cite{hamidouche2022versatile}, transcoder, cache management, and ABR algorithm, along with their other properties, such as required computation, bandwidth, and storage resources —key aspects that affect the performance and market competitiveness of consumer electronic devices.

\subsection{Control and Management Layer}
The Control and Management layer plays a central role in the REVISION architecture. Its main task is to enhance the performance of the streaming service regarding the specified objectives and constraints in the above layer and available resources in the lower layer, referred to as the Application and Resource layers, respectively. This layer's strategies are critical for developing consumer devices that maintain high-quality streaming under varied network conditions and usage scenarios.
The Control and Management Layer comprises a Centralized Controller and Optimizer (CCO), which serves as the core component by coordinating two controller modules: Resource Controller and Application Controller. The two controllers provide valuable information from their respective layers to support CCO in optimizing the streaming service. In the following, the components are introduced in detail:
\subsubsection{CCO Component--}
The CCO is the central intelligence of our video streaming architecture, designed to provide a seamless streaming experience across all types of consumer devices. It operates at the intersection of data provided by the Application and Resource Controller modules, employing advanced algorithms and strategies to optimize various aspects of the streaming experience. The CCO consists of three main modules: \textit{Coordinator}, \textit{Optimizer}, and \textit{Scheduler}. 
The Coordinator module is the main entity communicating with other controller modules, gathering information from them, and sending Optimizer commands to them. For example, if the Optimizer determines the need for additional resources for a streaming service, the Coordinator module communicates with the Resource Controller and requests the necessary adjustments.
The Optimizer module is responsible for carrying out optimization tasks, using inputs from the Coordinator, which are derived from the controller modules. In video streaming, where numerous events occur every second (\eg generated segments, player requests, bandwidth fluctuations, start/finish of streaming sessions), it is impractical to react to each event individually. Instead, a common approach is to define a time slot during which the Optimizer and Coordinator perform their tasks.
A typical time slot is divided into two intervals: data collection and optimization. During the data collection interval, monitoring parameters and events are gathered, and the Optimizer then executes its procedures and conveys the results to the Coordinator during the optimization interval. The duration of the time slot is highly dependent on the type of application. Shorter time slots impose greater computational demands and overhead on the system, while longer time slots may diminish the quality of results and the agility of the system~\cite{tashtarian_artemis_2023}.

\subsubsection{Application Controller--}
To provide useful information from the application layer, we use the Application Controller module that periodically monitors the provided streaming applications for consumer electronics to analyze their behaviors and deliver essential data for the CCO. By considering the aforementioned tasks, we can divide the operations of the Sub-Controller into two sub-modules: 
\begin{itemize}
    \item {Application Monitoring.}
    Before launching a video streaming application, the streaming service provider must define the constraints of the application. Then, during the streaming service, the Application Controller should track and assess various aspects of the streaming application's performance and user experience. This includes: 
    ($i$) Quality of Service (QoS) monitoring: checks video quality (\eg resolution, bit rate, frame rate, audio) for expected standards, ($ii$) Quality of Experience (QoE) monitoring: monitors buffering, playback interruptions, and latency for a smooth experience, ($iii$) error and issue detection: identifies and logs streaming errors (\eg error codes, playback failures), ($iv$) bandwidth and resource usage by the streaming functions: monitors bandwidth and resources for efficient operation, ($v$) user engagement metrics: gathers user data, popular content, and engagement for recommendations, ($vi$) security and compliance monitoring: ensures security and legal compliance, safeguarding the service. 
    \item {Application Analyzer.}
    The Application Analyzer is designed to obtain useful information from data gathered by the Application Monitoring module. Its primary responsibility is to meticulously analyze these data to extract meaningful insights and share them with the CCO. The Application Analyzer module serves the following main functions: ($i$) data interpretation: it transforms complex data from Application Monitoring into actionable insights, allowing real-time performance assessment, ($ii$) performance analysis: identifies areas for improvement by analyzing QoS metrics, user experience data, and error logs, ($iii$) user behavior insights: analyzes user behavior data to understand viewer preferences and content engagement, aiding recommendation systems, and ($iv$) quality assurance: ensures that the streaming function consistently meets or exceeds quality standards by tracking video and audio parameters.
\end{itemize}
An example of industry-based video streaming Application Controllers is Bitmovin Analytics\footnote{Bitmovin Analytics: \url{https://bitmovin.com/video-analytics/}} that continuously monitors video streaming functions, such as the encoder and player, to improve the overall user experience, content offerings, and business strategies of a streaming service. The Application Controller module, as shown in Figure~\ref{fig:arch}, plays a crucial role in making data-driven decisions for the ongoing success and optimization of the platform.
\subsubsection{Resource Controller--}
The Resource Controller's primary responsibilities are monitoring resource usage, allocating resources efficiently, and intelligently assigning tasks to ensure that the platform operates appropriately during the streaming service. This means optimizing device performance to maintain responsiveness even under heavy streaming loads in consumer electronics. As shown in Figure~\ref{fig:arch}, the Resource Controller includes the following three sub-modules:
\begin{itemize}
\item {Resource Monitoring}. The Resource Monitoring sub-module continuously tracks the availability and utilization of essential resources (\ie computation, storage, and bandwidth) in the Resource layer. It also evaluates the health and performance of resources, identifying any anomalies or potential issues. This includes monitoring for resource constraints, bottlenecks, and fluctuations in resource availability.
\item {Resource Allocation}. The Resource Allocation sub-module is responsible for distributing the available resources to the various video streaming functions according to the given solution by the CCO module. It allocates computing power, network bandwidth, and storage capacity based on the real-time requirements of the application layer. Moreover, based on the CCO output, it should dynamically scale the resources to accommodate fluctuations in demand, ensuring that resources are allocated optimally during peak usage periods and also scale down during lower-demand periods.
\item {Task Assignment}. Regarding the solutions determined by the CCO, the Task Assignment sub-module should accurately assign tasks, \ie encoding, transcoding, and data processing, to available resources. Task assignment can adaptively allocate tasks based on changing conditions in the Resource layer (\eg resource failures) and application constraints, ensuring tasks perform efficiently.
\end{itemize}

\subsection{Resource Layer}
This layer provides critical information about the available resources that can be used for video streaming applications. This is particularly relevant to devices with limited processing power, where efficient computation resource management can prolong device life and enhance performance. As shown in Figure~\ref{fig:arch}, the Resource Type encompasses a variety of resources essential for streaming services, including computation, storage, and bandwidth. Computation resources are primarily used in the content provisioning processes, such as encoding content into different segments. Moreover, when employing transcoding functions at the network edge, a significant amount of computation is required~\cite{erfanian2021oscar}.
Storage resources are critical in content distribution, where cache servers at the network edge store content to reduce serving latency. Additionally, players allocate a portion of their storage capacity as a buffer for storing downloaded segments. While this buffer size is relatively small compared to the storage capacity of CDNs, it significantly impacts the perceived QoE by players.
The last resource type is bandwidth, which presents a challenge in optimizing video streaming due to its fluctuations in all parts of the video streaming pipeline. Moreover, demand for substantial bandwidth is one of the most significant challenges in some video streaming applications such as immersive media~\cite{sai2023consumer}.
Since having up-to-date information about the current state of resources, including the available resources and their costs, is crucial for the CCO component, the Resource layer employs \textit{Resource Cost and Availability} module to inform valuable information to the Resource Sub-Controller regularly.\\

\section{Related Work}\par
Most proposed models for optimizing video streaming performance are based on Software Defined Networking (SDN) ~\cite{tashtarian2018s2vc, erfanian2020optimizing, bentaleb2017sdnhas, kleinrouweler2016delivering}. SDN separates the control plane (which makes decisions about where traffic should be sent) from the data plane (which forwards the actual network traffic). This separation allows researchers to utilize a centralized network management to configure and manage video streaming networks, optimize traffic flow, and respond to changing network requirements and conditions. 

In~\cite{tashtarian2018s2vc}, we introduce an SDN-based framework called S2VC, which aims to enhance QoE metrics and fairness in Scalable Video Coding (SVC)-based HTTP adaptive streaming. S2VC utilizes SDN controllers to optimize adaptation and data paths for delivering video files from servers to clients. 
We extended our work by proposing a real-time video multicast framework using SDN and Network Function Virtualization (NFV). We employ VRPs and Virtual Transcoder Functions (VTFs) to efficiently manage video streaming, reducing bandwidth usage and path selection effort. 
Bentaleb~\etal~\cite{bentaleb2017sdnhas} propose SDNHAS an intelligent streaming architecture that uses SDN to help adaptive streaming players make better decisions. It can handle large-scale deployments, reduce communication overhead, and allocate network resources effectively, even when the network experiences changes. Similarly, a SDN-based network architecture is proposed in~\cite{kleinrouweler2016delivering}. It offers two adaptation assistance mechanisms: signaling bitrates to DASH players and dynamic traffic control. The study demonstrates that this architecture enhances the streaming experience by doubling video bitrate and minimizing disruptions, providing valuable insights for implementing DASH-aware networking for various stakeholders.
Huang~\etal in~\cite{Digital2023xhyang} design a network architecture called DTN4VS (Digital Twin-driven Network for Video Streaming), with the aim of enabling network virtualization and customized network administration. Using this architecture, different kinds of Digital Twins (DTs) can describe the status of physical entities, separate network management tasks from the network controller, and enhance these functions with simulated data and customized strategies. To improve network management efficiency, they suggest three potential strategies: exploiting domain-specific data, evaluating performance, and adaptively updating the DT model.

In~\cite{borcoci2010novel}, a layered none-SDN architecture based on Content-Aware Networking (CAN) and Network-Aware Applications (NAA), is proposed with a focus on the virtual CAN layer. This architecture is part of a European FP7 ICT research project called ALICANTE and aims to meet the requirements for multimedia distribution and services across multiple IP domains while also addressing the needs of the Future Internet. It introduces a flexible business model and outlines the role of the CAN layer, its requirements, and interfaces with other system layers. The architecture offers QoS capabilities but also faces challenges, particularly related to the performance and cost of CAN devices in real network environments.

\section{CONCLUSION}\par
In this paper, we presented REVISION, a comprehensive roadmap for optimizing adaptive video streaming. The ever-increasing demand for online video content has made HTTP Adaptive Streaming (HAS) a pivotal technology for ensuring a high-quality viewing experience, particularly in consumer electronics such as mobile devices, smart TVs, and gaming consoles. REVISION focuses on improving video streaming by considering three main components: the Objective, the Input Space, and the Action Domain. By carefully balancing factors like QoE, latency, and cost, it provides a structured approach to address the challenges in video streaming, making it highly relevant for developers and manufacturers in the consumer technology sector.
Furthermore, the REVISION architecture introduces three layers: the Application Layer, the Control and Management Layer, and the Resource Layer. These layers work together to streamline the optimization process for video streaming, monitoring and adjusting various parameters and resources to meet the desired objectives and constraints. With the continued growth of video streaming, REVISION provides a valuable framework to enhance the viewing experience while managing the complexities of the evolving streaming ecosystem. This is crucial for consumer electronics companies aiming to deliver superior products that meet the dynamic needs of today’s consumers.

\section{ACKNOWLEDGMENTS}
The financial support of the Austrian Federal Ministry for Digital and Economic Affairs, the National Foundation for Research, Technology and Development, and the Christian Doppler Research Association are gratefully acknowledged. Christian Doppler Laboratory ATHENA: https://athena.itec.aau.at/

{\bibliography{main}}

\begin{thebibliography}{10}

\bibitem{nightingale2014video}
James Nightingale, Qi~Wang, Christos Grecos, and Sergio Goma.
\newblock Video adaptation for consumer devices: opportunities and challenges offered by new standards.
\newblock {\em IEEE Communications Magazine}, 52(12):157--163, 2014.

\bibitem{hamidouche2022versatile}
Wassim Hamidouche, Thibaud Biatek, Mohsen Abdoli, Edouard Fran{\c{c}}ois, Fernando Pescador, Milo{\v{s}} Radosavljevi{\'c}, Daniel Menard, and Mickael Raulet.
\newblock Versatile video coding standard: A review from coding tools to consumers deployment.
\newblock {\em IEEE Consumer Electronics Magazine}, 11(5):10--24, 2022.

\bibitem{khani2023recl}
Mehrdad Khani, Ganesh Ananthanarayanan, Kevin Hsieh, Junchen Jiang, Ravi Netravali, Yuanchao Shu, Mohammad Alizadeh, and Victor Bahl.
\newblock $\{$RECL$\}$: Responsive $\{$Resource-Efficient$\}$ continuous learning for video analytics.
\newblock In {\em 20th USENIX Symposium on Networked Systems Design and Implementation (NSDI 23)}, pages 917--932, 2023.

\bibitem{mattioli2020history}
Michael Mattioli.
\newblock History of video game distribution.
\newblock {\em IEEE Consumer Electronics Magazine}, 10(2):59--63, 2020.

\bibitem{chen2023understanding}
Hailiang Chen, Yifan Dou, and Yongbo Xiao.
\newblock Understanding the role of live streamers in live-streaming e-commerce.
\newblock {\em Electronic commerce research and applications}, 59:101266, 2023.

\bibitem{cisco2020cisco}
U~Cisco.
\newblock Cisco annual internet report (2018--2023) white paper.
\newblock {\em Cisco: San Jose, CA, USA}, 10(1):1--35, 2020.

\bibitem{noor2015connected}
Ahmed~K Noor.
\newblock The connected life.
\newblock {\em Mechanical engineering}, 137(09):36--41, 2015.

\bibitem{bentaleb2017sdnhas}
Abdelhak Bentaleb, Ali~C Begen, Roger Zimmermann, and Saad Harous.
\newblock {SDNHAS: An SDN-enabled architecture to optimize QoE in HTTP adaptive streaming}.
\newblock {\em IEEE Transactions on Multimedia}, 19(10):2136--2151, 2017.

\bibitem{tashtarian2022hxl3}
Farzad Tashtarian, Abdelhak Bentaleb, Alireza Erfanian, Hermann Hellwagner, Christian Timmerer, and Roger Zimmermann.
\newblock {HxL3: Optimized delivery architecture for HTTP low-latency live streaming}.
\newblock {\em IEEE Transactions on Multimedia}, 2022.

\bibitem{peroni2023empowerment}
Leonardo Peroni, Sergey Gorinsky, Farzad Tashtarian, and Christian Timmerer.
\newblock Empowerment of atypical viewers via low-effort personalized modeling of video streaming quality.
\newblock {\em Proceedings of the ACM on Networking}, 1(CoNEXT3):1--27, 2023.

\bibitem{huang2019comyco}
Tianchi Huang, Chao Zhou, Rui-Xiao Zhang, Chenglei Wu, Xin Yao, and Lifeng Sun.
\newblock Comyco: Quality-aware adaptive video streaming via imitation learning.
\newblock In {\em Proceedings of the 27th ACM international conference on multimedia}, pages 429--437, 2019.

\bibitem{lyko2024improving}
Tomasz Lyko, Matthew Broadbent, Nicholas Race, Mike Nilsson, Paul Farrow, and Steve Appleby.
\newblock Improving quality of experience in adaptive low latency live streaming.
\newblock {\em Multimedia Tools and Applications}, 83(6):15957--15983, 2024.

\bibitem{zhang2021performance}
Bo~Zhang, Thiago Teixeira, and Yuriy Reznik.
\newblock Performance of low-latency http-based streaming players.
\newblock In {\em Proceedings of the 12th ACM Multimedia Systems Conference}, pages 356--362, 2021.

\bibitem{tashtarian2018s2vc}
Farzad Tashtarian, Alireza Erfanian, and Amir Varasteh.
\newblock {S2VC: An SDN-based framework for maximizing QoE in SVC-based HTTP adaptive streaming}.
\newblock {\em Computer Networks}, 146:33--46, 2018.

\bibitem{erfanian2020optimizing}
Alireza Erfanian, Farzad Tashtarian, Reza Farahani, Christian Timmerer, and Hermann Hellwagner.
\newblock {On optimizing resource utilization in AVC-based real-time video streaming}.
\newblock In {\em 2020 6th IEEE Conference on Network Softwarization (NetSoft)}, pages 301--309. IEEE, 2020.

\bibitem{iraschko2019next}
Rainer Iraschko and James Slevinsky.
\newblock Next-generation video network design tenets: Building a better video delivery service.
\newblock {\em IEEE Consumer Electronics Magazine}, 8(4):22--27, 2019.

\bibitem{rahman2023content}
Waqas~ur Rahman and Eui-Nam Huh.
\newblock Content-aware qoe optimization in mec-assisted mobile video streaming.
\newblock {\em Multimedia Tools and Applications}, 82(27):42053--42085, 2023.

\bibitem{tashtarian_artemis_2023}
Farzad Tashtarian, Abdelhak Bentaleb, Hadi Amirpour, Sergey Gorinsky, Junchen Jiang, Hermann Hellwagner, and Christian Timmerer.
\newblock {ARTEMIS: Adaptive Bitrate Ladder Optimization for Live Video Streaming}.
\newblock In {\em 21st USENIX Symposium on Networked Systems Design and Implementation (NSDI)}, 2023.

\bibitem{tashtarian_lalisa_2023}
Farzad Tashtarian, Abdelhak Bentaleb, Hadi Amirpour, Babak Taraghi, Christian Timmerer, Hermann Hellwagner, and Roger Zimmermann.
\newblock {LALISA: Adaptive Bitrate Ladder Optimization in HTTP-Based Adaptive Live Streaming}.
\newblock In {\em IEEE/IFIP Network Operations and Management Symposium (NOMS)}, 2023.

\bibitem{erfanian2021lwte}
Alireza Erfanian, Hadi Amirpour, Farzad Tashtarian, Christian Timmerer, and Hermann Hellwagner.
\newblock {LwTE: light-weight transcoding at the edge}.
\newblock {\em IEEE Access}, 9:112276--112289, 2021.

\bibitem{erfanian2021oscar}
Alireza Erfanian, Farzad Tashtarian, Anatoliy Zabrovskiy, Christian Timmerer, and Hermann Hellwagner.
\newblock {OSCAR: On optimizing resource utilization in live video streaming}.
\newblock {\em IEEE Transactions on Network and Service Management}, 18(1):552--569, 2021.

\bibitem{erfanian2021lwte-live}
Alireza Erfanian, Hadi Amirpour, Farzad Tashtarian, Christian Timmerer, and Hermann Hellwagner.
\newblock {Lwte-live: Light-weight transcoding at the edge for live streaming}.
\newblock In {\em Proceedings of the Workshop on Design, Deployment, and Evaluation of Network-Assisted Video Streaming}, pages 22--28, 2021.

\bibitem{nguyen2023performance}
Minh Nguyen, Hadi Amirpour, Farzad Tashtarian, Christian Timmerer, and Hermann Hellwagner.
\newblock Performance analysis of h2br: Http/2-based segment upgrading to improve the qoe in has.
\newblock {\em Multimedia Tools and Applications}, pages 1--35, 2023.

\bibitem{mao2017neural}
Hongzi Mao, Ravi Netravali, and Mohammad Alizadeh.
\newblock Neural adaptive video streaming with pensieve.
\newblock In {\em Proceedings of the conference of the ACM special interest group on data communication}, pages 197--210, 2017.

\bibitem{afzal2022otec}
Samira Afzal, Farzad Tashtarian, Hamid Hadian, Alireza Erfanian, Christian Timmerer, and Radu Prodan.
\newblock {OTEC: an optimized transcoding task scheduler for cloud and fog environments}.
\newblock In {\em Proceedings of the 2nd International Workshop on Design, Deployment, and Evaluation of Network-Assisted Video Streaming}, pages 21--26, 2022.

\bibitem{bhat2021combining}
Madhukar Bhat, Jean-Marc Thiesse, and Patrick Le~Callet.
\newblock Combining video quality metrics to select perceptually accurate resolution in a wide quality range: A case study.
\newblock In {\em 2021 IEEE International Conference on Image Processing (ICIP)}, pages 2164--2168. IEEE, 2021.

\bibitem{licciardello2022prepare}
Melissa Licciardello, Lukas Humbel, Fabian Rohr, Maximilian Gr{\"u}ner, and Ankit Singla.
\newblock Prepare your video for streaming with segue.
\newblock {\em Journal of Systems Research (JSys)}, 2(1), 2022.

\bibitem{turkkan2022greenabr}
Bekir~Oguzhan Turkkan, Ting Dai, Adithya Raman, Tevfik Kosar, Changyou Chen, Muhammed~Fatih Bulut, Jaroslaw Zola, and Daby Sow.
\newblock Greenabr: energy-aware adaptive bitrate streaming with deep reinforcement learning.
\newblock In {\em Proceedings of the 13th ACM Multimedia Systems Conference}, pages 150--163, 2022.

\bibitem{lorenzi2024medusa}
Daniele Lorenzi, Farzad Tashtarian, Hermann Hellwagner, and Christian Timmerer.
\newblock Medusa: A dynamic codec switching approach in http adaptive streaming.
\newblock {\em ACM Transactions on Multimedia Computing, Communications and Applications}, 2024.

\bibitem{sai2023consumer}
Siva Sai, Dishank Goyal, Vinay Chamola, and Biplab Sikdar.
\newblock Consumer electronics technologies for enabling an immersive metaverse experience.
\newblock {\em IEEE Consumer Electronics Magazine}, 2023.

\bibitem{kleinrouweler2016delivering}
Jan~Willem Kleinrouweler, Sergio Cabrero, and Pablo Cesar.
\newblock {Delivering stable high-quality video: An SDN architecture with DASH assisting network elements}.
\newblock In {\em Proceedings of the 7th International Conference on Multimedia Systems}, pages 1--10, 2016.

\bibitem{Digital2023xhyang}
Xinyu Huang, Haojun Yang, Shisheng Hu, and Xuemin Shen.
\newblock Digital twin-driven network architecture for video streaming.
\newblock {\em IEEE Network}, 2024.

\bibitem{borcoci2010novel}
Eugen Borcoci, Daniel Negru, and Christian Timmerer.
\newblock {A novel architecture for multimedia distribution based on content-aware networking}.
\newblock In {\em 2010 Third International Conference on Communication Theory, Reliability, and Quality of Service}, pages 162--168. IEEE, 2010.

\end{thebibliography}

\end{document}